\documentclass[a4,12pt]{article}
\usepackage[dvips]{graphicx}
\usepackage{amsmath,amssymb}
\newcommand{\abs}[1]{\left\vert#1\right\vert}
\date{}
\begin{document}

\title{\bf Testing a theory of gravity in celestial mechanics:
a new method and\\
its first results for a scalar theory}

\author{M. Arminjon}
\maketitle
\footnote{~This is a revised and expanded version of the text of a lecture
given at the $8^{\mathrm{th}}$ Conf. ``Physical Interpretations of
Relativity Theory'' (London, September 2002), organized by the
British Society for the Philosophy of Sciences. The initial version will appear in the Proceedings of that Conference (M. C. Duffy, ed.).
} 
\begin{center} { \small Laboratoire ``Sols, Solides, Structures'' [Unit\'e Mixte de Recherche of the CNRS], 
\\ BP 53, F-38041 Grenoble cedex 9, France}
\\ email: \verb+arminjon@hmg.inpg.fr+

\vspace{2cm}

\begin{abstract} 
A new method of post-Newtonian approximation (PNA) for weak
gravitational fields is presented together with its application to
test an alternative, scalar theory of gravitation. The new method
consists in defining a one-parameter family of systems, by
applying a Newtonian similarity transformation to the initial data
that defines the system of interest. This method is rigorous. Its
difference with the standard PNA is emphasized. In particular, the
new method predicts that the internal structure of the bodies does
have an influence on the motion of the mass centers. The translational equations of motion 
obtained with this method in the scalar theory are adjusted in the solar
system, and compared with an ephemeris based on the standard PNA of GR.
\end{abstract}

\end{center}
\maketitle
\section{Introduction}
One first expects from a theory of gravity that it should provide an accurate
celestial mechanics. In other words, the theory should tell us how
massive celestial bodies precisely move with respect to each other
under the effect of the gravitational field produced by them all.
Thus, Einstein's general relativity (GR) won its first advantage
over the older theory of Newton when it gave an explanation to
Mercury's residual advance in perihelion. In 1972, Weinberg stated
about this explanation (\cite{Weinberg72}, p.~198):``This is by far the most
important experimental verification of general relativity." It is
hence extremely important for a theory of gravitation, not only
that it produces accurate ephemerides, but even more that one is
sure that it really produces {\it those} ephemerides, {\it i.e.}, that
the solution of the approximate equations used in the computation
does approach accurately enough the relevant solution of the exact
equations. The works of Fock \cite{Fock59} and Chandrasekhar \cite{Chandra65} aimed at
answering the latter question for GR. The later work on
celestial mechanics in GR relies on essentially the same approximation scheme as these
two works, which are equivalent in this regard. Yet in 1966, Synge, who did know Fock's work
(which is quoted in Ref. \cite{Florides62}) and most probably knew also that of Chandrasekhar,
wrote \cite{Synge66}:``I am still waiting for a rational treatment of the
dynamics of the solar system according to Einstein's theory. In
the very nature of the case, any argument must be of an
approximate nature; an assessment of the error is a primary
desideratum." Comparing his successive sentences, we may infer
that Synge was not satisfied with the approximation method used in
the works \cite{Fock59, Chandra65} nor with the one he himself proposed with
coworkers \cite{Das61, Florides62}, and which was limited to stationary fields---this restriction is indeed inappropriate to describe the solar system in a realistic way.\\

The aim of this paper is to summarize the principles, the
development and the numerical implementation of a new
approximation method for celestial mechanics in relativistic
theories of gravitation. This approximation method might have
satisfied Synge, perhaps, because it is mathematically sound and
general, and because it too predicts a salient result which he
found in his work for GR \cite{Das61, Florides62}, namely the fact that, in such
theories, the internal structure of a body does influence the
gravitational field produced by it, hence also the motion of
external bodies \cite{Arm00b2, Arm02b}. The new method consists basically in associating a one-parameter family $(\mathrm{S}_\lambda)$ of gravitating systems with the physically given system $\mathrm{S}$, by defining a family of initial conditions. It was initiated by Futamase \& Schutz \cite{FutaSchutz} for GR, with further mathematical developments given
by Rendall \cite{Rendall92}. However, Futamase \& Schutz~\cite{FutaSchutz} assumed a very restrictive initial condition for the spatial metric. As to Rendall \cite{Rendall92}, he considered an a priori given one-parameter family of solutions of the field equations, without investigating the definition of a such family from the given system $\mathrm{S}$. Moreover, these two works were limited to the local equations and some of their mathematical properties. In
particular, they did not provide equations of motion for the mass
centers of a system of extended bodies, as one needs to compute an
ephemeris. We came to the new method independently \cite{Arm98a,Arm00a}, to
test an alternative theory of gravitation based on just a scalar
field \cite{Arm97, Arm96}, and we did obtain such equations of motion \cite{Arm00b1, Arm00b2, Arm02b}. That scalar theory gives the same predictions as GR for light
rays \cite{Arm98a}. Therefore, it is worth testing this theory further.
Moreover, since that theory is much simpler than GR, it is easier
to implement the new method for that theory, as well as to discuss
the difference between the new method and the standard PNA.

\section{General framework: the method of asymptotic
expansions}

As is well-known, an asymptotic expansion of a real function
$\varphi$ of the real variable $\lambda$ in the neighborhood of
some value $\Lambda$ is an expression
\begin{equation} \label{philambda}
    \varphi(\lambda)=a_{0}\psi_{0}(\lambda)+...+a_{n}\psi_{n}(\lambda)+R(\lambda),
\end{equation}
the known functions $\psi_{0}$, ..., $\psi_{n}$ being positive and
belonging to a definite {\it comparison set} E, endowed with
certain properties, and with  $\psi_{0}\gg ... \gg\psi_{n}\gg R$
as $\lambda\rightarrow\Lambda$ \cite{Dieudonne68}. In physics, the relevant
value is usually $\Lambda = 0$, and one speaks of the ``small
parameter" $\lambda$. In particular, if the behaviour as
$\lambda\rightarrow0$ is regular enough, a Taylor expansion may
apply, so that $\psi_{k}(\lambda)=\lambda^k (k = 0,..., n)$.
However, it may be that the Taylor expansion can be pushed only to
some order $n=n_{\mathrm{max}}$, beyond which a more accurate
expansion can be obtained only if one accepts to consider more
general functions, {\it e.g.} ones involving a fractional exponent $k$.
This remark is relevant to weak-field expansions in relativistic
theories of gravitation.\\

    Now consider a boundary-value problem defined for a given system
of partial differential equations (PDE's), and assume that a small
parameter $\lambda$ can be defined for this problem, which means
in fact that a {\it family} $(\mathrm{P}_{\lambda})$ of problems
can be defined. The method of asymptotic expansions for this
problem consists in trying to write {\it each scalar component} of
the solution of $\mathrm{P}_{\lambda}$, say
$\varphi_{i}^{(\lambda)}$ (where $i=1,...,m$ with $m$ the number
of scalar unknowns involved in the system of PDE's), as an asymptotic expansion in $\lambda$. That expansion is thus assumed valid {\it at each given point} $X\in\mathrm{D}$, where
$\mathrm{D}$ is the relevant domain for the independent variables
(space and time, say), which are collectively denoted by $X$.
Actually this domain itself may well depend on $\lambda$, but, in
order that one may write definite expansions (\ref{philambda})
involving known functions $\psi_{k}(\lambda)$, it is preferable to
absorb this dependence in a redefinition of the independent
variables such that the expansions indeed apply to any given point
$X$ in a domain independent of $\lambda$. If we look for a Taylor
expansion, we write thus:
\begin{equation}\label{philambdaX}
  \forall X\in \mathrm{D},\quad \varphi_{i}^{(\lambda)}(X)=\varphi_{i0}(X)+...+ 
  \varphi_{in}(X)\lambda^n +R_i (X,\lambda),
\end{equation}
with $R_i (X,\lambda)\ll\lambda^n$ as $\lambda\rightarrow0$. The
problem which is really of physical interest, $\mathrm{P}$, {\it e.g.}
the initial-value problem for some (assumed) isolated
self-gravitating system, is assumed to correspond to a given,
small value $\lambda_0$ of the parameter. One of the difficulties
of the method is definition of an adequate family
$(\mathrm{P}_{\lambda})$, from the given problem $\mathrm{P}$.
Once this has been done and once expansions as
$\lambda\rightarrow0$, and corresponding expanded equations, have
been obtained for the family $(\mathrm{P}_{\lambda})$, they are
then used for the finite value $\lambda_0$. This means that the
error involved is the value for $\lambda_0$ of the unknown
remainder $R_i (X,\lambda)$. In the case of a Taylor expansion,
however, the remainder will usually be $O(\lambda^{n+1})$, and
this uniformly with respect to $X\in \mathrm{D}$ (the relevant
domain $\mathrm{D}$ being often compact). Although this is only an
{\it asymptotic} error estimate, thus not a {\it numerical} one,
it can be said that, if $\lambda^{n+1}$ is negligible with respect
to the experimental accuracy, then the $n$-th order expansion
(\ref{philambdaX}) is very likely to be enough accurate for a
meaningful experimental test. This seems to be the best that one
can hope in the current state of the theory of PDE's.\\

Why is it useful to write expansions at all? Mainly because the
resulting equations are very greatly simplified: as it turns out,
all non-linearities are reported in the equations of the order
zero, and those are often much simpler than the starting
equations. We emphasize two points:\\

i) It is indeed necessary to define a {\it family} of
boundary-value problems (instead of contenting oneself with just
that problem which one is interested in). This is in order that it
just {\it make sense} to try expansions in $\lambda$: if we have
defined a such family, we can then expand with respect to
$\lambda$ each of the various coefficients that define the system
of PDE's, and we also can expand the boundary values. Only in that
case can we derive expansions of the solution fields, and expanded
equations for them, and then solve these equations using the
expanded boundary conditions.\\

ii) One should indeed expand {\it all} independent unknown fields,
$\varphi_{i}^{(\lambda)}$ for $i=1,...,m$ (and not just those
which one likes to expand). For this allows one to write each
equation as a sum of terms, each of a definite order in $\lambda$,
this allowing in turn to separate the equations of the different
orders: $k=0,...,n$---which is necessary, because an
expansion like (\ref{philambdaX}) means that a field
$\varphi_{i}^{(\lambda)}$ is, after expansion, split into the
$(n+1)$ fields $\varphi_{i0},...,\varphi_{in}$. Thus, if we write
expansions for the $m$ independent unknown fields, using
expansions with $(n+1)$ terms ({\it e.g.} Taylor expansions of order
$n$), then we get each of the $m$ independent equations split into
$(n+1)$ equations, so that we now have $m(n+1)$ equations for
$m(n+1)$ unknowns. Whereas, if one expands only $p$ among the
independent fields (with $p<m$), using expansions with $(n+1)$
equations, then one has (of course!) no reason to split the
equations. But if one nevertheless would do so, then one would have too much equations (if $n\geq
1$): $m(n+1)$ equations for $p(n+1)+m-p$ unknowns.\\

The above-described method is merely the general formulation of a
natural perturbation method for a system of PDE's, and it is
certainly not new. Of course there is a vast literature on
perturbation methods (see {\it e.g.} Refs. 17--18 and references
therein), and there are many common points between the different
approaches. Yet we have not been able to find a description like
the foregoing one, which applies very closely to what we actually
did for the scalar theory of gravitation. Certainly also, the
knowledge of PDE's and of rigorous perturbation methods has
considerably improved since the fifties. This may explain why the
method developed for weak fields in GR by Fock \cite{Fock59} and
Chandrasekhar \cite{Chandra65} does not fulfil the requirements i) and ii)
above---in fact it is based on formally taking $1/c^2$ as a small
parameter (with $c$ the velocity of light), and in expanding the
gravitational field, but not the matter fields, in powers of
$1/c^2$.

\section{Asymptotic post-Newtonian approximation of the scalar
theory}

It is the application of the foregoing method to the case of that
``relativistic'' theory of gravitation.
\footnote{~The scalar
theory investigated is indeed relativistic in the sense that it
accounts for special relativity, and reduces to it if the
gravitational constant $G$ vanishes---but it is a preferred-frame
theory. The summary of the scalar theory that is given in Ref.
\cite{Arm00a}, Sect. 2, is sufficient (and not even necessary, we
believe) for the present purpose. Being based on a scalar field,
that theory is, of course, very different from the {\it
relativistic theory of gravitation} proposed by Logunov {\it et
al.} \cite{Logunov88,Logunov89}. However, both theories consider a
flat ``background'' metric and a curved ``effective'' metric.
} 
In such theories, the relevant boundary-value problem is the {\it
initial-value problem}. This is due to the hyperbolic character of
the gravitational equation, in other words it comes from the fact
that, in such theories, gravitation propagates with a finite
velocity, usually equal to the velocity of light $c$ or close to
$c$. This applies \cite{Arm00a} to the scalar theory investigated by the
author. Note, however, that the complete system of local equations
is not closed, hence in particular cannot be qualified
``hyperbolic", until one has postulated a definite behaviour for
matter, by assuming a constitutive equation giving the material
energy-momentum tensor $\mathbf{T}$ in terms of some matter
fields. For the sake of simplicity, we assume a perfect barotropic
fluid, for which one has only the pressure $p$ and the (spatial)
velocity $\mathbf{u}$ as independent matter fields.
\footnote
{~The nature of the relevant boundary-value problem might {\it a priori} be
expected to depend on the constitutive equation assumed. However,
even if one assumed a very general matter behaviour, including
anelasticity and dissipation effects, it seems that the
initial-value problem would remain the ``good problem". The reason
to believe this is that it is so for the heat equation, although
the latter is parabolic.}\\

Thus we consider a given, isolated gravitating system
$\mathrm{S}$, made of $N$ separated bodies. That system is defined
by the barotropic state equations in the different bodies, and by
the initial data for the independent matter fields $p$ and
$\mathbf{u}$, as well as for the scalar gravitational field $f$
and for its time derivative $\partial _{T}f$. (In GR, an initial
data for the gravitational field is much more complicated, because the
latter field is a tensor one, also because the Einstein
equations are underdetermined, and above all because the initial conditions have to verify nonlinear 
constraint equations \cite{Stephani82,Rendall00}.) The
first task is to define a family $(\mathrm{S}_{\lambda})$ of
gravitating systems, by defining a family of initial data and
state equations. We want to describe weakly gravitating systems,
for which Newton's theory with its Euclidean space and absolute
time must be an excellent approximation. For this to be true, the
family  $(\mathrm{S}_{\lambda})$ must satisfy two conditions as
$\lambda\rightarrow0$: i) the (physical, or ``effective")
space-time metric $\boldsymbol{\gamma}$ must tend towards a flat
metric $\boldsymbol{\gamma}^0$, and ii) all fields must become
closer and closer to ``corresponding" Newtonian fields. For the
scalar theory, which is bimetric, the flat metric
$\boldsymbol{\gamma}^0$ always coexists with the physical one
$\boldsymbol{\gamma}$, and their local difference is an increasing
function of $1-f$, which is non-negative \cite{Arm00a}. Condition i) is
hence easy to be made precise for the scalar theory: it means
simply that the maximum value of the field $(1-f^{(\lambda)})$
must tend towards zero as $\lambda\rightarrow0$. The parameter is
\begin{equation}\label{deflambda}
  \lambda= \mathrm{Sup}_{\mathbf{x}\in \mathrm{M}}[1-f^{(\lambda)}(\mathbf{x},T=0)]/2.
\end{equation}
where $\mathrm{M}$ is the ``space" manifold, {\it i.e.} the set of the
positions $\mathbf{x}$ in the preferred reference frame (PRF).\\

To make condition ii) precise, we use a crucial result \cite{FutaSchutz, Arm00a},
namely the existence of an {\it exact similarity transformation in
Newton's theory} for barotropic fluids. Suppose we have the
following fields: pressure $p^{(1)}$, density
$\rho^{(1)}=F^{(1)}(p^{(1)})$, Newtonian potential $U_{N} ^{(1)}$,
and velocity $\mathbf{u}^{(1)}$, obeying the continuity equation,
Poisson's equation, and Euler's equation. Then, for any
$\lambda>0$, the fields $p^{(\lambda)}(\mathbf{x},T)=\lambda^2
p^{(1)}(\mathbf{x},\sqrt{\lambda}\ T)$,
$\rho^{(\lambda)}(\mathbf{x},T)=\lambda
\rho^{(1)}(\mathbf{x},\sqrt{\lambda}\ T)$,
$U_{N}^{(\lambda)}(\mathbf{x},T)=\lambda
U_{N}^{(1)}(\mathbf{x},\sqrt{\lambda}\ T)$, and
$\mathbf{u}^{(\lambda)}(\mathbf{x},T)=\sqrt{\lambda}\
\mathbf{u}^{(1)}(\mathbf{x},\sqrt{\lambda}\ T)$, also obey these
equations---provided the state equation for system
$\mathrm{S}_\lambda$ is $F^{(\lambda)}(p^{(\lambda)}) = \lambda
F^{(1)}(\lambda^{-2} p^{(\lambda)})$. This similarity
transformation defines the weak-field limit in Newton's theory
itself: as $\lambda\rightarrow0$, the potential and the density in
the bodies decrease like $\lambda$ (while the bodies keep the same
size), the (orbital) velocities decrease like $\sqrt{\lambda}$,
and accordingly the time scale increases like $1/\sqrt{\lambda}$.\\

This exact similarity transformation does not extend to a
``relativistic" theory like the scalar theory, simply because the
equations are different. In particular, the metric changes as the
gravitational field becomes stronger, and this changes the motion
in a complex manner. Recall, however, that we merely seek {\it
initial conditions} for the fields in the scalar theory. It is
then obvious that, in order to precise condition ii), we simply
have to {\it apply the Newtonian similarity transformation to the
initial fields}. For the matter fields, this application is indeed
immediate, because for a perfect fluid these are common to
Newton's theory and to a relativistic theory (via some slight
modifications): pressure $p$, proper rest-mass density $\rho^*$
(instead of the invariant density of Newton's theory), mass
density of elastic energy $\Pi$, coordinate velocity
$\mathbf{u}=d\mathbf{x}/dT$ (for the scalar theory, $T$ is the
preferred time coordinate or ``absolute time", and $\mathbf{x}$ is
the position in the preferred reference frame (PRF)). The mere
difficulty is that, to apply the transformation, we must also
associate a kind of ``Newtonian potential" with the scalar
gravitational field $f$. To do that, we assume in advance that a
family $(\mathrm{S}_\lambda)$ of systems has been built, which is
such that the orders in $\lambda$ of the matter fields are the
same, as $\lambda\rightarrow0$, as in the Newtonian similarity
transformation, and such that the field $f^{(\lambda)}$ admits an
expansion of the form $f^{(\lambda)}=1+\phi \lambda^{k} +
O(\lambda^{k+1})$. We find then that one must have $k=1$, {\it i.e.} the
field $1-f^{(\lambda)}$ is like $\lambda$ as
$\lambda\rightarrow0$, thus like the potential in the Newtonian
weak-field limit.\footnote{~The small parameter considered in the
present paper is $\lambda=\varepsilon^2$, where $\varepsilon$  is
that used in Ref. 12.} It is hence natural to define that field,
more precisely $V^{(\lambda)} = (c^2/2)(1-f^{(\lambda)})$, as the
equivalent of the Newtonian potential $U_N^{(\lambda)}$; the
coefficient is needed to ensure that $V^{(\lambda)}\sim
U_N^{(\lambda)}$ as $\lambda\rightarrow0$. Moreover, the local
difference between the flat metric $\boldsymbol{\gamma}^0$ and the
curved physical metric $\boldsymbol{\gamma}$ is an increasing
function of $1-f^{(\lambda)}(\mathbf{x},T$).\\

In that way, we get the initial conditions for system
$\mathrm{S}_\lambda$. Then we state powers expansions in $\lambda$
for the independent fields $f$, $p$, $\mathbf{u}$, and we deduce
expansions for the other fields. The set of the obtained
expansions and expanded equations is internally consistent,
moreover it is consistent with the initial aim, in that the
equations of the order 0 are indeed the ``Euler-Newton" equations,
{\it i.e.} the equations for a perfect fluid in Newton's theory. The
first PN corrections correspond to the equations of the order 1.
The details can be found in Ref. 12 (see Ref. 15 for a synopsis
and a few complementary points). Let us mention here some
important points:\\

-- We use a change of the mass and time units for system
$\mathrm{S}_\lambda$, adopting $[\mathrm{M}]_\lambda =
\lambda[\mathrm{M}]$ and $[\mathrm{T}]_\lambda =
[\mathrm{T}]/\sqrt{\lambda}$  as the new units (where
$[\mathrm{M}]$ and $[\mathrm{T}]$ are the starting units); then
all fields are $\mathrm{ord}(\lambda^0)$, and the small parameter
$\lambda$ is proportional to $1/c^2$ (in fact
$\lambda=(c_{0}/c)^2$, where $c_0$ is the velocity of light in the
starting units). That $1/c^2$, not $1/c$, turns out to be the
effective small parameter, is due to the fact that it is only
$1/c^2$ that enters the equations. Thus, the derivation of the
expansions and expanded equations is straightforward. In the
standard PN scheme \cite{Fock59, Chandra65}, $1/c^2$ is {\it formally} considered as
a small parameter, and the matter fields are not expanded.\\

-- Recall, however, that such Taylor expansions in $\lambda$ (or
$1/c^2$) cannot be expected to apply to an arbitrary order $n$
[see after Eq.~(\ref{philambda})]. In GR, it has been shown by
Rendall \cite{Rendall92} that the attainable order $n$ depends on the gauge
condition: with the classical ``harmonic gauge", even the second
PN approximation ($n = 2$) cannot be solved for a general
spatially compact system of bodies---but a different gauge
condition allows to reach the order $n = 3$. This kind of problem
cannot occur in the scalar theory, because there is no gauge
condition. However, only the first PN approximation ($n = 1$) has
been studied in the scalar theory. If the matter sources are
considered given (as was done in GR by Rendall \cite{Rendall92}), then the PN
gravitational field depends on integrals whose integrand is
regular and vanishes outside the bodies [see Eqs.
(\ref{f-expans})--(\ref{W}) below], so the first PN approximation
is unproblematic.\\

-- The equations of the order 1 are linear with respect to the
fields of the order 1, so that the nonlinearities are confined to
the zero-order (Euler-Newton) equations. The separation between
the matter fields of the different orders, as well as the
linearity in the PN fields, are characteristic of the asymptotic
method, as contrasted with the standard PN scheme \cite{Fock59, Chandra65}. It seems
worth to illustrate the difference between the two approximations
by exhibiting some simple equations.

\section{Illustrating and commenting the difference between
``standard" and ``asymptotic" PNA's}

In the asymptotic PNA of the scalar theory, the rest-mass density
in the PRF, $\rho_\mathrm{exact}$, is approximated
as\footnote{~The asymptotic expansion has the form
(\ref{rhoexact})$_1$ in the ``varying units" defined above, but it
has a slightly different form in invariable units \cite{Arm00a}.}
\begin{equation}\label{rhoexact}
  \rho_\mathrm{exact}=\rho_{(1)}[1+ O(\lambda^2)],\qquad \rho_{(1)}\equiv \rho+\rho_{1}/c^2,
\end{equation}
and the like for the other fields, {\it e.g.} the velocity field
$\mathbf{u}_{\mathrm{exact}}$. (For the zero-order, Newtonian
fields, we shall omit the index 0 and shall thus keep the usual
notation, while denoting henceforth the exact fields by the
subscript ``exact".) The first-order (1PN) expansion of the time
component of the local equations of motion for a perfect fluid is
\cite{Arm00a}
\begin{equation}\label{T-ord0}
  \partial _T \rho + \partial _j (\rho u^j) = 0,
\end{equation}
\begin{equation}\label{T-ord1}
  \partial _T (w+\rho_1) + \partial _j [(w+p+\rho_1)u^j +\rho
  u_1^{j}]=-\rho \, \partial_T U,\qquad w\equiv\rho(\frac{\mathbf{u}^2}{2}+\Pi -U)
\end{equation}
($U\equiv \mathrm{N.P.}[\rho]$ is the Newtonian potential
associated with the 0-order mass density $\rho$). The 0-order
spatial component of the local equations of motion is just the
Newtonian equation of motion:
\begin{equation}\label{i-ord0}
    \partial _T (\rho u^i) + \partial _j (\rho u^i u^j)=\rho
    U_{,i}-p_{,i}.
\end{equation}
Combining (\ref{i-ord0}), the continuity equation (\ref{T-ord0}),
and the 0-order expansion of the isentropy equation:
\begin{equation}\label{isentropy}
  d\Pi = -p \, d(1/\rho),
\end{equation}
one gets in a standard way the Newtonian energy equation:
\begin{equation}\label{energyN}
  \partial _T w + \partial _j [(w+p)u^j]=-\rho \, \partial_T U.
\end{equation}
Subtracting (\ref{energyN}) from (\ref{T-ord1}) gives us
\begin{equation}\label{mass-ord1}
  \partial _T \rho_1 + \partial _j(\rho_1 u^j + \rho u_1 ^j)=0,
\end{equation}
which means that mass is conserved at the first PNA of the scalar
theory.\\

In contrast, the standard 1PN expansion of the time component of
the local equations of motion for a perfect fluid in GR consists
(\cite{Chandra65}, Eqs.~(1) and (64)) of Eq.~(\ref{T-ord0}), plus
\begin{equation}\label{Chandra-T-PN}
    \partial _T \sigma +  \partial _j (\sigma u^j)+ \frac{1}{c^2}
    (\rho \partial _T U- \partial _T p) =0,\quad \sigma\equiv
    \rho[1+\frac{1}{c^2}(\mathbf{u}^2 +2U+\Pi+\frac{p}{\rho})].
\end{equation}
(Fock \cite{Fock59} gives only an integral form of the equations of motion.)
In the equations~(\ref{T-ord0}) and~(\ref{Chandra-T-PN}) of the
standard PNA of GR, however, $\rho$, $\mathbf{u}$ and $p$ are not
defined except as exact fields (\cite{Chandra65}, Eqs.~(4) and~(5)), and,
actually, they have to be considered as the second approximation
of the exact fields---instead of being the zero-order expanded
fields as in Eqs.~(\ref{T-ord0}) and (\ref{T-ord1}) of the
asymptotic PNA of the scalar theory. Thus, each of the matter
fields is there only in one ``exemplar" in the equations of the
standard PNA. On the other hand, Chandrasekhar \cite{Chandra65} does not regard
Eqs.~(\ref{T-ord0}) and~(\ref{Chandra-T-PN}) as exact ones:
rather, Eq.~(\ref{T-ord0}) is assumed valid if one neglects
$O(1/c^2)$ terms, and Eq.~(\ref{Chandra-T-PN}) is assumed valid if
one neglects $O(1/c^4)$ terms.
\footnote{~This is clear, {\it e.g.}, from
the sentence after Eq.~(109) in Ref.~\cite{Chandra65} (which is just the
repetition of Eq.~(\ref{Chandra-T-PN}) above, Eq.~(64) in
Ref.~\cite{Chandra65}): ``That [equation] replaces, in the post-Newtonian
approximation, the equation of continuity of Newtonian
hydrodynamics. We may transform the terms in equation~(109) that
occur explicitly with the factor $1/c^2$ with the aid of the
equations valid in the Newtonian limit."
} 
Also note that, unlike
our Eq.~(\ref{T-ord1}) of the asymptotic PNA, the PN
equation~(\ref{Chandra-T-PN}) is not a ``split" equation. Just the
same remarks apply to the spatial components of the local
equations of motion. As to the gravitational field, it is expanded
in the standard PNA: at the first approximation, there is only the
Newtonian potential $U$, whereas ``PN potentials" $U_i$ and $\Phi$
are added in the second approximation. The Poisson equation
applies to $U$:
\begin{equation}\label{Poisson}
  \Delta U=-4\pi G \rho
\end{equation}
(Eq.~(3) in Ref.~\cite{Chandra65}, Eq.~(68.25) in Ref.~\cite{Fock59}), and Poisson-like
equations are derived for $U_i$ and $\Phi$.\\

In our opinion, the difficulty with the standard PNA is this:
since the matter fields are not expanded, it follows that the
equations of the first approximation are not exact ones resulting
from a ``splitting" (as it is the case in the asymptotic PNA).
Instead, they result from a truncation and can, indeed, be valid
only if one neglects $O(1/c^2)$ terms (as Fock and Chandrasekhar
noted). This point has been {\it proved} for the Poisson equation
(\ref{Poisson}) in the scalar theory, in Ref. \cite{Arm00a}, Sect. 6.2. Thus, the ``asymptotically correct" writing of Eq.~(\ref{Poisson}) of the standard PNA is
\begin{equation}\label{PoissonOK}
  \Delta U=-4\pi G \rho +O\left(\frac{1}{c^2}\right).
\end{equation}

But this means that the field $U$ (the Newtonian potential) can be considered to be
known only if one neglects $O(1/c^2)$ terms. Now that field
provides the main contribution to the acceleration, hence the
equation of motion of the second approximation (Eq.~(68) in
Ref.~3) does of course include a term involving that field $U$
{\it without} an $1/c^2$ factor, namely the term $\rho \partial
_iU$. This means that, at least if one restricts the discussion to
the {\it local} equations, the second approximation does not
provide a better approximation than up to $O(1/c^2)$ terms not
included---that is, it does not actually improve over the first
approximation. Thus, if the first standard PNA is to really reach
the $1/c^2$ level {\it included}, it must turn out that, at the
later stage of the global equations of motion for the mass
centers, and due to some mysterious integration effect, the low
accuracy to which the Poisson equation can be considered valid has
no effect any more. Moreover, as noted by Rendall \cite{Rendall92}, it
may be dangerous to have PDE's which are defined only up to
unknown higher-order terms, which can change the type of the equation, {\it e.g.} from hyperbolic to elliptic. In conclusion, we prefer to stay with
the ``asymptotic" PNA summarized above,
and to see where it leads.

\section{Equations of motion for the mass centers: general PN
equations}

One may first ask whether there may be any satisfying {\it
definition} of the mass center of a body in a ``relativistic"
theory of gravitation, given that: (i) in a theory accounting for
the mass-energy equivalence, any form of material energy should be
{\it subjected to} the action of the gravitational field; (ii)
conversely, any form of material energy should {\it contribute to}
the gravitational field; this means that the internal structure
(through its energy distribution) and the internal motion (through
the corresponding kinetic energy) are {\it a priori} expected to play a
role~\cite{Arm00b1}; for these two reasons, it is not obvious at
all whether one single scalar mass-energy density can be used so
as to define a single mass center, and which density this should
be ~\cite{Arm00b1}; (iii) in a generally-covariant theory like GR,
a covariant definition should be found for the mass center---in
other words, the motion of the latter should correspond to a
single world-line in space-time, independently of which reference
frame has been chosen~\cite{Tucker02}. The latter difficulty does
not exist in the scalar theory, which has a preferred reference
frame, but the two first ones subsist. Moreover, one has to ensure
that the definition adopted will be {\it astronomically relevant}.
With modern telescopes and other instruments, the major bodies of
the solar system are, of course, very-well resolved as extended
bodies, hence the astronomical position refers to an ``optical
center". It seems to be a good approximation to admit that, once corrected from the ``phase effects", the latter corresponds to averaging the rest-mass density, because it is the body's
``matter", in the usual sense, that does radiate electromagnetic
energy. In any case, this density is certainly more relevant than
any density involving gravitational energy, for the latter is
distributed in the whole space. Therefore, we define the (exact) mass
center by averaging the (exact) rest-mass density in the preferred frame,
$\rho_{\mathrm{exact}}$ \cite{Arm00b1}. A theoretical argument
also favours that density, and this is the fact that its PN
approximation $\rho_{(1)}$, at least, obeys the usual continuity
equation (see Eqs.~(\ref{T-ord0}) and (\ref{mass-ord1})), thus
without adding gravitational energy and its flux: this implies
that the velocity of the mass center is itself the barycenter of
the velocity---when the barycenter is defined with $\rho_{(1)}$
as the weight function~\cite{Arm00b1}.\\

Thus, we define the exact mass and mass center through the
rest-mass density $\rho_{\mathrm{exact}}$ :
\begin{equation}\label{defmasscent}
  M_a ^{\mathrm{exact}}\equiv\int_{\mathrm{D}_a}\rho_{\mathrm{exact}}dV,\qquad M_a ^{\mathrm{exact}}\mathbf{a}_{\mathrm{exact}}\equiv\int_{\mathrm{D}_a}\rho_{\mathrm{exact}}\mathbf{x}dV
\end{equation}
where $\mathrm{D}_a$ is the (time-dependent) domain occupied by
body $(a)$ ($a=1,..., N$), in the PRF ($V$ is the Euclidean volume
measure in the PRF). At the (first) PNA, the mass and the mass
center are approximated by
\footnote{~The notation might suggest that, in Eq. (\ref{defPNmasscent}), the $M_a^1 \mathbf{a}_1/c^2$ term is order 2, as the product of two first-order terms $M_a^1$
and $\mathbf{a}_1$ (still multiplied by the constant coefficient
$1/c^2$). Recall, however, that the expansions are written in the
varying units $[\mathrm{M}]_\lambda = \lambda[\mathrm{M}]$ and
$[\mathrm{T}]_\lambda = [\mathrm{T}]/\sqrt{\lambda}$. In these
units, $M_a^1$ and $\mathbf{a}_1$ are coefficients that do not
depend of $\lambda$, thus are order 0, while $1/c^2$ is
proportional to $\lambda$. Hence the $M_a^1 \mathbf{a}_1/c^2$ term
is really order 1, not 2. Similarly, in Eq. (\ref{masscent-ord1}),
all terms are order zero. If one wishes, one may come back to the
starting units (independent of $\lambda$), in which the orders of
the different fields are not all the same, and in which the
expansions are hence different, {\it e.g.}
$\rho_{\mathrm{exact}}=\lambda[\rho +\lambda
\rho_1/c^2]+O(\lambda^3)$ instead of (\ref{rhoexact}), with $\rho$
and $\rho_1$ still of order 0 (Ref. \cite{Arm00a}, Sect. 6). This complicates the
analysis of the orders, but of course it cannot lead to any
inconsistency. Anyway, in practice, we are then using these
expansions for one single value of the parameter $\lambda$, namely
the value $\lambda_0$ corresponding to the gravitating system of
physical interest (the solar system, say). Thus we can use the
expressions valid ``in the varying units", {\it e.g.} (\ref{rhoexact}).
}
\begin{equation}\label{defPNmass}
  M_a^{(1)}=M_a+M_a^1/c^2,\qquad M_a\equiv\int_{\mathrm{D}_a}\rho
  dV,\qquad M_a^1\equiv\int_{\mathrm{D}_a}\rho_1 dV,
\end{equation}
\begin{equation}\label{defPNmasscent}
  M_a^{(1)}\mathbf{a}_{(1)}\equiv\int_{\mathrm{D}_a}\rho_{(1)}\mathbf{x}dV=
  M_a\mathbf{a}+M_a^{1}\mathbf{a}_{1}/c^2,
\end{equation}
with
\begin{equation}\label{defmasscent-ord0-ord1}
  M_a\mathbf{a}=\int_{\mathrm{D}_a}\rho\mathbf{x}dV,\qquad M_a^{1}\mathbf{a}_{1}=\int_{\mathrm{D}_a}\rho_{1}\mathbf{x}dV
\end{equation}
Note that $M_a$ and $\mathbf{a}$ are the Newtonian mass and mass
center. Using Eqs.~(\ref{T-ord0}) and (\ref{mass-ord1}), one shows
that $M_a$ and $M_a^1$ are constant in time \cite{Arm00b1}. To get the PN
equations of motion of the mass centers, one just integrates the
spatial components of the PN local equations of motion inside the
different bodies \cite{Arm00b1}. Due to the separation of the different
orders in the local equations and to their linearity with respect
to the PN fields, separate equations are also obtained for the
mass centers, and the equation for PN corrections (order 1) is
linear with respect to order-1 quantities. Specifically one finds
\cite{Arm00b1}:
\begin{equation}\label{masscent-ord0}
  M_a\ddot{a}^i=\int_{\mathrm{D}_a} \rho U^{(a)}_{,i} dV
\end{equation}
and
\begin{equation}\label{masscent-ord1}
  M_a^1\,\ddot{a}_1^i+\dot{I}^{ai}=J^{ai}+K^{ai},
\end{equation}
with
\begin{equation}\label{Iai}
  I^{ai}\equiv\int_{\mathrm{D}_a} [p+\rho(\mathbf{u}^2/2+\Pi+U)]u^i dV,
\end{equation}
\begin{equation}\label{Jai}
  J^{ai}\equiv\int_{\mathrm{D}_a} (\sigma_1 U_{,i}+\rho A_{,i}) dV,
\end{equation}
\begin{equation}\label{Kai}
  K^{ai}\equivò\int_{\mathrm{D}_a}[2\mathsf{k}_{ij}p_{,j}+pU_{,i}
  -2U\rho U_{,i}-_1\Gamma_{jk}^i\rho u^ju^k-\rho u^j\partial_T \mathsf{k}_{ij}]dV.
\end{equation}
In these equations, a point means (total) derivative with respect
to the preferred time $T$; $\sigma_1$ is the PN correction to the
active mass density, given by
\begin{equation}\label{sigma1}
  \sigma_1 = \rho_1 + \rho (\mathbf{u}^2/2+ \Pi + U);
\end{equation}
$A$ is the PN gravitational potential, such that the PN expansion
of the scalar field is
\begin{equation}\label{f-expans}
  f^{(\lambda)} = 1-2 U/c^2-2 A/c^4 + O(\lambda^3),
\end{equation}
and A is given by:
\begin{equation}\label{A}
  A = B +  \partial  ^2 W/ \partial T^2,
\end{equation}
where $B \equiv \mathrm{N.P.}[\sigma_1]$ is the Newtonian
potential associated with $\sigma_1$ and
\begin{equation}\label{W}
  W(\mathbf{X},T) \equiv \int G R \rho(\mathbf{x},T)dV(\mathbf{x})/2
  \qquad R \equiv  \abs{\mathbf{X - x}}
\end{equation}
($G$ is the constant of gravitation); $2\mathsf{k}_{ij}/c^2$ and
$_1\Gamma_{jk}^i/c^2$ are the components of the non-Euclidean part
of the "physical" space metric in the PRF and its associated
connection, respectively, with 
\begin {equation} \label{Gammaijk}
_1\Gamma_{jk}^i \equiv \mathsf{k}_{ij,k} + \mathsf{k}_{ik,j} - \mathsf{k}_{jk,i}, 
\quad \mathsf{k}_{ij} \equiv U \mathsf{h}^1_{ij}, \quad \mathsf{h}^1_{ij} \equiv
\frac{U_{,i}U_{,j}}{U_{,k}U_{,k}};
\end{equation}
\{we use Cartesian coordinates for the Euclidean space metric
$\boldsymbol{g}^0$, so that the PN physical space metric writes \cite{Arm00a}
\begin{equation}\label{PNmetric}
  g^{(1)}_{ij} = \delta_{ij} + 2\mathsf{k}_{ij}/c^2 ;\}
\end{equation}
and we use Fock's decomposition~\cite{Fock59} of any field
$Z(\mathbf{x})$, integral of some density $\theta_\mathbf{x}$
vanishing outside the bodies, into ``self " and ``external" parts
$z_a(\mathbf{x})$ and $Z^{(a)}(\mathbf{x})$ :
\begin{equation}\label{self-external}
  Z(\mathbf{x})\equiv\int\theta_\mathbf{x} dV = z_a(\mathbf{x})
+Z^{(a)}(\mathbf{x}),\quad z_a(\mathbf{x})
\equiv\int_{\mathrm{D}_{a}}\theta_\mathbf{x} dV,
Z^{(a)}(\mathbf{x}) \equiv\sum_{b\neq
a}\int_{\mathrm{D}_{b}}\theta_\mathbf{x}dV.
\end{equation}
The zero-order equation, Eq.~(\ref{masscent-ord0}), is just the
Newtonian translational equation of motion.

\section{Equations of motion for the mass centers: relevant
simplifications}

Equation (\ref{masscent-ord1}) for the PN corrections to the
motion is not tractable. To make it explicit, we account for two
simplifications that do occur for the major bodies of the solar
system, namely: (i) the good separation between bodies (this
occurs probably also for many other systems); (ii) the fact that
the bodies are nearly spherical (this occurs almost certainly also
for all other systems, if one considers individual bodies with
large-enough mass).\\

Point (i) is defined by introducing the separation parameter \cite{Arm00b2}
\begin{equation}\label{eta}
  \eta_0\equiv\max_{a \neq b}(r_b/\abs{\mathbf{a-b}}),
  \qquad r_b\equiv \frac{1}{2}\mathrm{Sup}_{\mathbf{x},\mathbf{y}\in\mathrm{D}_b}\abs{\mathbf{x-y}},
\end{equation}
and by assuming that it is small for the system of physical
interest. (The zero-order positions of the mass centers are used,
since the PN corrections to these positions are very small, hence
would lead to nearly the same value for $\eta_0$.) In order to
exploit this small parameter, we introduce again a (conceptual)
family of gravitational systems, by defining initial conditions
for them \cite{Arm02b}. Thus, remembering the weak-field parameter
$\lambda$, we actually would have a two-parameters family of
systems. However, the local 1PN equations of the asymptotic method---thus the set of the first-order expansions, like (\ref{rhoexact}), the set of the expanded equations, like
(\ref{T-ord0}) and (\ref{mass-ord1}), and the set of the expanded
boundary conditions, derived in Refs.~12 and 15---are
mathematically self-consistent, in that they make a closed system
of equations. (They are not physically self-consistent yet, since
they do not ensure by themselves that the weak-field/low-velocity
assumptions are satisfied and remain so in time.) Hence we may
content ourselves with deducing a one-parameter family
$(\mathrm{S}'_\eta)$ of well-separated ``PN systems" from the data
of the given PN system $\mathrm{S}'$, the latter being itself
deduced from the given system $\mathrm{S}$ by substituting the 1PN
equations for the exact ones. (Actually the local PN equations of
the asymptotic method {\it are} exact, but they define merely a
part of the exact fields \cite{Arm00a}.) The PN gravitational fields: $U =
\mathrm{N.P.}[\rho]$, $B = \mathrm{N.P.}[\sigma_1]$, and $W$,
depend merely on the PN matter fields
[Eqs.~\ref{f-expans}--\ref{W}]. Hence we just have to define the
initial conditions for the PN matter fields. Moreover, the latter
conditions turn out to be determined by the initial zero-order
matter fields $p(T = 0)$, or equivalently $\rho(T = 0)$, and
$\mathbf{u}(T = 0)$ \cite{Arm00a}. To define $\rho(T = 0)$ in
$\mathrm{S}'_\eta$, we set
\begin{equation}\label{a-eta-T=0}
 \mathbf{a}^\eta(T = 0) = \mathbf{a}(T = 0) \eta_0/\eta,
\end{equation}
\begin{equation}\label{rho-eta-T=0}
  \rho^{\eta}(\mathbf{x},T=0)=\rho(\mathbf{a+y},T=0)\quad
  \mathrm{if}\quad\mathbf{x}= \mathbf{a}^\eta + \mathbf{y}\quad\mathrm{with}\quad\mathbf{a+y}\in \mathrm{D}_a.
\end{equation}
Equation (\ref{rho-eta-T=0}) defines $\rho^{\eta}$ so that the
density inside the bodies is independent of $\eta$ [setting
$\rho^{\eta}(\mathbf{x}, T = 0) = 0$ if $\mathbf{x}$ does not have
the form above for some $a = 1, ..., N$]. In other words, the
bodies themselves do not depend on the separation parameter
$\eta$. Equation (\ref{a-eta-T=0}) ensures that, at least near $T
= 0$, the separation distances between bodies are of order
$\eta^{-1}$:
\begin{equation}\label{a-eta-b-eta}
  \abs{\mathbf{a}^{\eta}-\mathbf{b}^{\eta}}=
  \mathrm{ord}(\eta^{-1}).
\end{equation}
To define the velocity $\mathbf{u}^\eta(T = 0)$, we use the
auxiliary assumption that each body undergoes a rigid motion at
the Newtonian approximation \cite{Fock59}:
\begin{equation}\label{u-rigid}
  u^i =  \dot{a}^i + \Omega^{(a)}_{ji}(x^j - a^j ),\,\mathrm{or}\,
  \mathbf{u} =\dot{\mathbf{a}}+ \boldsymbol{\omega}_{a}\wedge(\mathbf{x-a})
  \, \mathrm{for} \, \mathbf{x}\in\mathrm{D}_a,\,\Omega^{(a)}_{ji} +\Omega^{(a)}_{ij} = 0.
\end{equation}
Of course, this is only approximately true ({\it e.g.} due to the tidal
influence of the other bodies), but we use this assumption merely
to calculate the PN corrections. Since the latter ones are very
small, it certainly implies only an extremely small error in the
solar system. Anyhow, we can assume that (\ref{u-rigid}) is exact
at the initial time. From the Newtonian estimate
$\dot{\mathbf{a}}^2\approx U^{(a)}(\mathbf{a})$ , valid in the
reference frame of the global mass center, and from
(\ref{a-eta-b-eta}), we expect that, at any time,
\begin{equation}\label{adot-eta}
  (\dot{a}^i)^\eta=\mathrm{ord}(\eta^{1/2}).
\end{equation}
We assume that this is true if we define the initial translation
velocities of system $\mathrm{S}^\eta$  as
\begin{equation}\label{adot-eta-T=0}
  (\dot{a}^i)^\eta (T=0)=(\eta/\eta_0)^{1/2}\dot{a}^i(T=0).
\end{equation}
As to the self-rotation velocities, in the solar system they have
at most the same magnitude, in linear values, as the translation
velocities, and our numerical calculations show that the PN
corrections containing quadratic terms in $\Omega^{(a)}_{ji}$,
included in the first version of the explicit translational
equations \cite{Arm00b2}, are negligibly small in the solar system. To avoid
such terms in the expansions, it turns out to be sufficient that
\begin{equation}\label{omega-eta}
  (\Omega^{(a)}_{ji})^\eta\ll\eta^{1/2},
\end{equation}
hence we set, for some small number $\varepsilon > 0$,
\begin{equation}\label{omega-eta-T=0}
  (\Omega^{(a)}_{ji})^\eta
  (T=0)=(\eta/\eta_0)^{\varepsilon+1/2}\,\Omega^{(a)}_{ji}(T=0),
\end{equation}
and we assume that this ensures that (\ref{omega-eta}) is true at
any time.\\

Point (ii) is defined simply by assuming, merely at the stage of
calculating the PN {\it corrections}, that the zero-order
rest-mass density $\rho$ is spherically symmetric for each body:
\begin{equation}\label{spheric-rho}
  \forall \mathbf{x}\in\mathrm{D}_a,\,\rho(\mathbf{x})=\rho_a(r),
  \quad r\equiv\abs{\mathbf{x-a}}\quad (a = 1, ..., N).
\end{equation}
The sphericity of the field $\rho$ implies also that of the
pressure field $p$ and the Newtonian self-potential $u_a$. By
using this and point (i), {\it i.e.} Eqs.~(\ref{a-eta-b-eta}),
(\ref{adot-eta}) and (\ref{omega-eta}), we get after rather
involved calculations \cite{Arm00b2, Arm02b}:
\begin{eqnarray}\label{big-one}
  \frac{d\mathbf{u}_{1a}}{dT}& = & o(\eta^3)+
  \left[\left(\frac{\tau_a}{M_a}-\frac{5}{6}\right)\mathbf{u}_{a}^{2}-
  \frac{5}{3}U^{(a)}(\mathbf{x}_a)-\frac{17\varepsilon_a+6T_a}{3M_a}\right]\nabla
  U^{(a)}(\mathbf{x}_a)\nonumber\\
  & & -\left[\left(\frac{\tau_a}{M_a}+2\right)\mathbf{u}_{a}\mathbf{.}\nabla
  U^{(a)}(\mathbf{x}_a)\right]\mathbf{u}_{a}+\left[2\frac{\varepsilon_a}{M_a}+
  U^{(a)}(\mathbf{x}_a)\right]\boldsymbol{\omega}_{a}\wedge
  \mathbf{u}_a\nonumber\\
  & & +G\sum_{b\neq a}\left\langle\frac{M_b}{2r_{ab}}\left[\left(\mathbf{n}_{ab}\mathbf{.}
  \dot{\mathbf{u}}_b\right)\mathbf{n}_{ab}
  -\dot{\mathbf{u}}_b\right]\right.\nonumber\\
  & & +\frac{1}{r_{ab}^2}\left\{-\alpha'_{ab}\mathbf{n}_{ab}+
  \frac{M_b}{2}\left[\left(3(\mathbf{n}_{ab}\mathbf{.}\mathbf{u}_b)^2-
  \mathbf{u}_b^2\right)\mathbf{n}_{ab}
  -\left(\mathbf{n}_{ab}\mathbf{.}\mathbf{u}_b\right)\left(2\mathbf{u}_b+
  \frac{8}{3}\mathbf{u}_a\right)\right]\right\}\nonumber\\
  & & +\left.\frac{M_b}{r_{ab}^3}\left[\mathbf{x}_{1b}\mathbf{-}\mathbf{x}_{1a}+
  3\left((\mathbf{x}_{1a}\mathbf{-}\mathbf{x}_{1b})\mathbf{.}\mathbf{n}_{ab}
  \right)\mathbf{n}_{ab}\right]\right\rangle,
\end{eqnarray}
 with $r_{ab}\equiv\abs{\mathbf{x}_a-\mathbf{x}_b}$,
 $\mathbf{n}_{ab}\equiv(\mathbf{x}_a-\mathbf{x}_b)/r_{ab}$, and
\begin{equation}\label{alphaprime}
  \alpha'_{ab}\equiv M_b\left(\frac{\mathbf{u}_{a}^{2}+\mathbf{u}_{b}^{2}}{2}
  +U^{(a)}(\mathbf{x}_a)+U^{(b)}(\mathbf{x}_b)+\frac{11\varepsilon_a+8T_a}{3M_a}\right)
  +M_b^1+\frac{11}{3}\varepsilon_b+\frac{8}{3}T_b,
\end{equation}
and where, as a preparation for the application, the notation for
the PN positions and velocities of the mass centers has been
changed, as compared with (\ref{defPNmasscent}) and
(\ref{defmasscent-ord0-ord1}):
\begin{equation}\label{def-posi}
  \mathbf{x}_a \equiv \mathbf{a},\quad \mathbf{x}_{1a} \equiv
  c^2(\mathbf{a}_{(1)}-\mathbf{a}),\qquad
\mathbf{x}_{(1)a} \equiv \mathbf{a}_{(1)}= \mathbf{x}_a+
\mathbf{x}_{1a}/c^2
\end{equation}
\begin{equation}\label{def-velo}
  \mathbf{u}_a\equiv \dot{\mathbf{a}},\quad \mathbf{u}_{1a}
  \equiv c^2(\dot{\mathbf{a}}_{(1)}-\dot{\mathbf{a}})=\dot{\mathbf{x}}_{1a},
\quad\mathbf{u}_{(1)a}\equiv \dot{\mathbf{a}}_{(1)}= \mathbf{u}_a+
\mathbf{u}_{1a}/c^2
\end{equation}
Note that terms of order up to and including $\eta^3$ have been
consistently retained in Eq.~(\ref{big-one}) \cite{Arm02b}. In the first
version of the explicit translational equations of motion \cite{Arm00b2},
this could not be done due to the lack of an asymptotic framework
for the separation parameter $\eta$, and this resulted in a large
difference with observational data. In addition to the
self-rotational energy:
\begin{equation}\label{def-Ta}
  T_a\equiv \Omega^{(a)}_{ik}
  \Omega^{(a)}_{jk}I^{(a)}_{ij}/2,\qquad
   I^{(a)}_{ij}\equiv \int_{D_a} \rho(x^i-a^i)(x^j-a^j) dV ,
\end{equation}
 two {\it structure-dependent} parameters appear
in (\ref{big-one}):
\begin{equation}\label{def-epsa}
  \varepsilon_a\equiv\int_{D_a} \rho u_a dV/2,
\end{equation}
\begin{equation}\label{def-taua}
  \tau_a\equiv\frac{1}{3G} \int_0 ^{r_a} u_a \left\{\frac{4r\mu_a'}{\mu_a(r)}
  -\left[\frac{r\mu_a'}{\mu_a(r)}\right]^2\right\}dr, \qquad
\mu_a(r)\equiv 4\pi\int_0 ^r \rho_a(s) s^2 ds.
\end{equation}
No such structure parameter does enter in the PN equations that
have been used in celestial-mechanical tests of GR, namely the
Lorentz-Droste-Einstein-Infeld-Hoffmann (LDEIH) equations, which
are based on the standard PNA. However, as already noted by Synge
and coworkers, one should expect that the internal structure of the
bodies does influence the motion.

\section{Numerical implementation. Comparison with an ephemeris
based on the  standard PNA of GR}

In order to use the translational equations of motion
(\ref{masscent-ord0}) and (\ref{big-one}), so as to check the
scalar theory, we have to know the values of the independent
parameters that enter these equations. These are: the zero-order
masses $M_a$ of the bodies (here the major bodies of the solar
system) and their parameters $T_a$, $\varepsilon_a$ and $\tau_a$;
the initial conditions of their motion; and the constant velocity
$\mathbf{V}$ of the global zero-order mass center of the solar
system, with respect to the preferred frame E \cite{Arm00b2}. (Of course there
is no parameter like $\mathbf{V}$ in conventional theories.) These
unknown parameters depend on the theory. They have to be
determined by optimizing the agreement between predictions and
observations \cite{Arm00b2}. In particular, one may expect that the optimal
values of the zero-order parameters should differ from their
values in pure Newtonian theory by first-order
(second-approximation) quantities, this fact alone giving
corrections of the same magnitude as the 1PN corrections \cite{Arm00b2}.
However, since the parameters $T_a$, $\varepsilon_a$ and $\tau_a$
play a role only in the PN corrections, such first-order
corrections on their values would make only second-order
differences, hence negligible ones, in the final results.
Therefore, we calculate these parameters from the standard
rotation velocities and density profiles of the corresponding
bodies \cite{Bakouline75,Lewis95}.\\

Then our adjustment algorithm loops on the numerical solution of
the translational equations of motion in order to optimize the
remaining parameters, {\it i.e.}, the zero-order masses $M_a$, the
initial conditions, and the velocity $\mathbf{V}$. The algorithm
has been tested \cite{Arm02c} by investigating in which measure
one may reproduce (over one century) the predictions of the DE403
ephemeris \cite{Standish95}, by using purely Newtonian equations
of motion. (The DE403 ephemeris is based on LDEIH-type equations
of motion \cite{Newhall83}, thus it is based on the standard PNA
of GR.) Our algorithm has also been applied to adjust a less
simplified model, in which the PN corrections in the Schwarzschild
field of the Sun are also considered, and to compare this model
with DE406 \cite{Standish98} over 60 centuries \cite{Arm02a}. It
has thus been found that the difference between a calculation
based on these ``Schwarzschild-corrected" Newtonian equations,
limited to the Sun and the nine major planets, and the DE406
ephemeris of the JPL, based on LDEIH-type equations and including
the Moon and asteroids, is very small. Over the last century, for
instance, the longitude difference for Mercury is $0.03''$
\cite{Arm02a}. The VSOP82 ephemeris \cite{Bretagnon82} is also
based on ``Schwarzschild-corrected" Newtonian equations, but it
uses more accurate (semi-analytical) algorithms than ours; also,
for the comparison with the JPL ephemerides, it seems that it
takes Schwarzschild's metric in ``isotropic" form, rather than in
the standard form in which we took it. The VSOP82 ephemeris leads
to even smaller differences with an ephemeris based on LDEIH-type
equations ({\it e.g.} $0.001''$ over 1891-2000 for the longitude of
Mercury). Now the scalar theory does predict Schwarzschild's
motion for test particles in the field of one spherical body, if
the latter is at rest in the preferred frame ($\mathbf{V} = 0$).
Hence, if we take for PN corrections in our theory those that are
obtained by considering the planets as test particles in the field
of the spherical Sun, and if we include $\mathbf{V}$ in the free
parameters, then we can obtain only a {\it smaller} difference
with the LDEIH-type equations than the difference between the
latter and the ``Schwarzschild-corrected" Newtonian equations.
But, in our opinion, the correct equations of motion are those got
with the asymptotic PNA. Therefore, we have implemented the
equations (\ref{masscent-ord0}) and (\ref{big-one}) in our
adjustment algorithm.\\

To do that, we had first to solve a numerical shortcoming of the
asymptotic method, namely the fact that the mass center of a body,
or the test particle, is followed on its trajectory by two
different positions: the zero-order position $\mathbf{x}_0$ and
the 1PN position, $\mathbf{x}_{(1)}\equiv \mathbf{x}_0 +
\mathbf{x}_1/c^2$, which drift from one another as the time goes.
By investigating in detail the case of a test particle in a
Schwarzschild field, it has been found that this leads to an error
increase in $(T-T_0)^2$, and that it may be cured by
``reinitializing", {\it i.e.}, by substituting $T_0+ \delta T$, then
$T_0+ 2\delta T$, etc., with $\abs{\delta T}$ sufficiently small,
for the initial time $T_0$ in the differential system that governs
the motion of the mass centers \cite{Arm01}. Moreover, since the
equation for PN corrections (\ref{big-one}) is valid only in the
preferred reference frame E, we use a Lorentz transform to go from
E to the frame $\mathrm{E}_\mathbf{V}$ bound with the zero-order
global barycenter, and vice-versa. This transform, as  well  as
the inverse transform, is determined by the free vector
$\mathbf{V}$. Thus, the adjustment process of the translational
equations on observational data provides us eventually with the
value of $\mathbf{V}$ that minimizes the residual with the set of
observations. However, the ``observational data" are currently
taken from an {\it ephemeris} based on GR, specifically here we
took from DE403 a set of heliocentric positions of the eight major
planets (Pluto omitted), between 1956 and 2000.\\
\begin{figure}\label{1a4}
	\includegraphics[width=\textwidth]{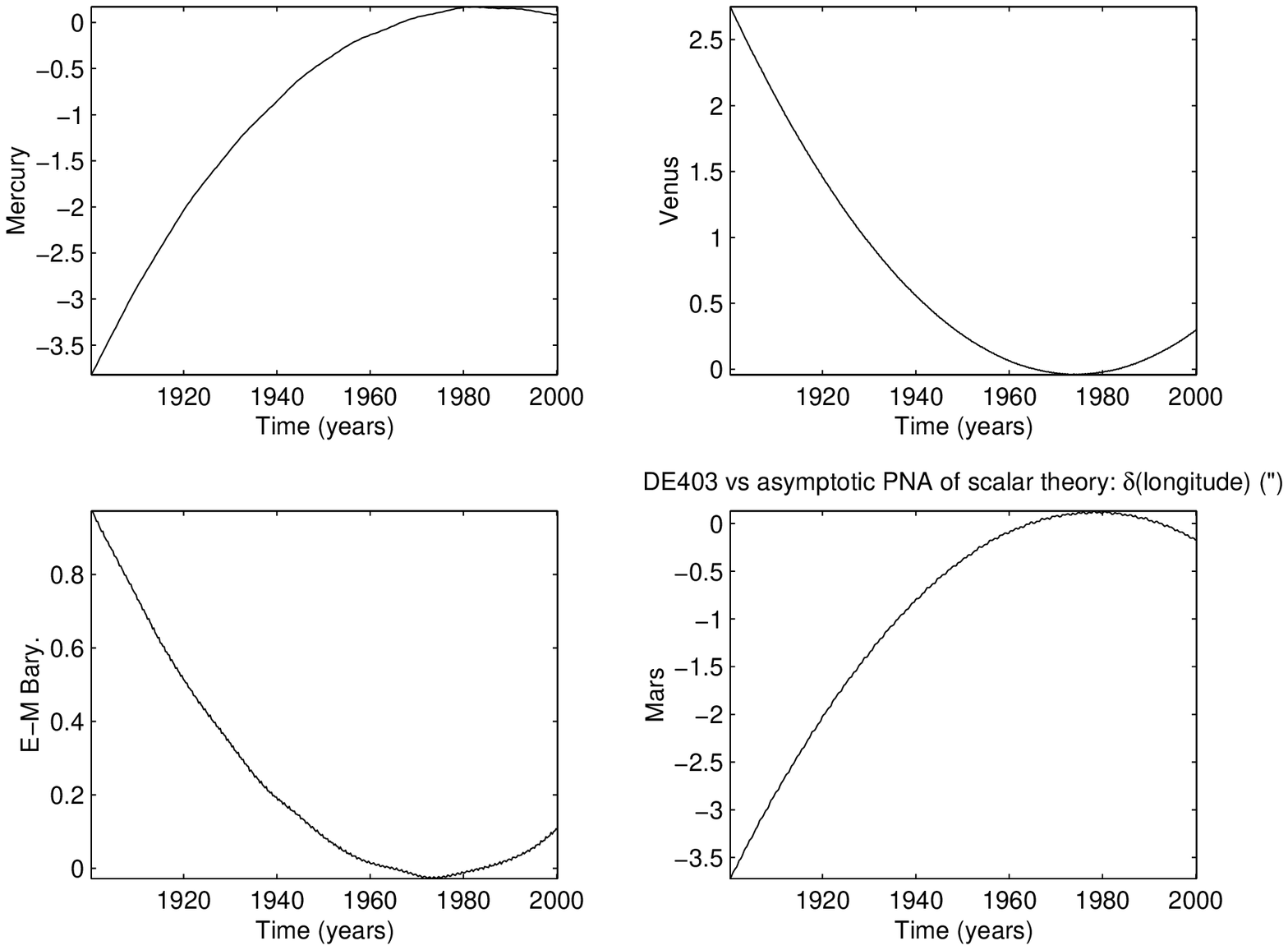}
	\caption{\small{Longitude differences (arc seconds) for 
	planets Mercury to March, either as taken from the DE403 ephemeris of the Jet
	Propulsion Laboratory, or as obtained (after adjustment) by
	numerical integration of the equations of motion in the asymptotic
	post-Newtonian approximation of the investigated scalar theory.}}
\end{figure} 
\begin{figure}\label{5a8}
	\includegraphics[width=\textwidth]{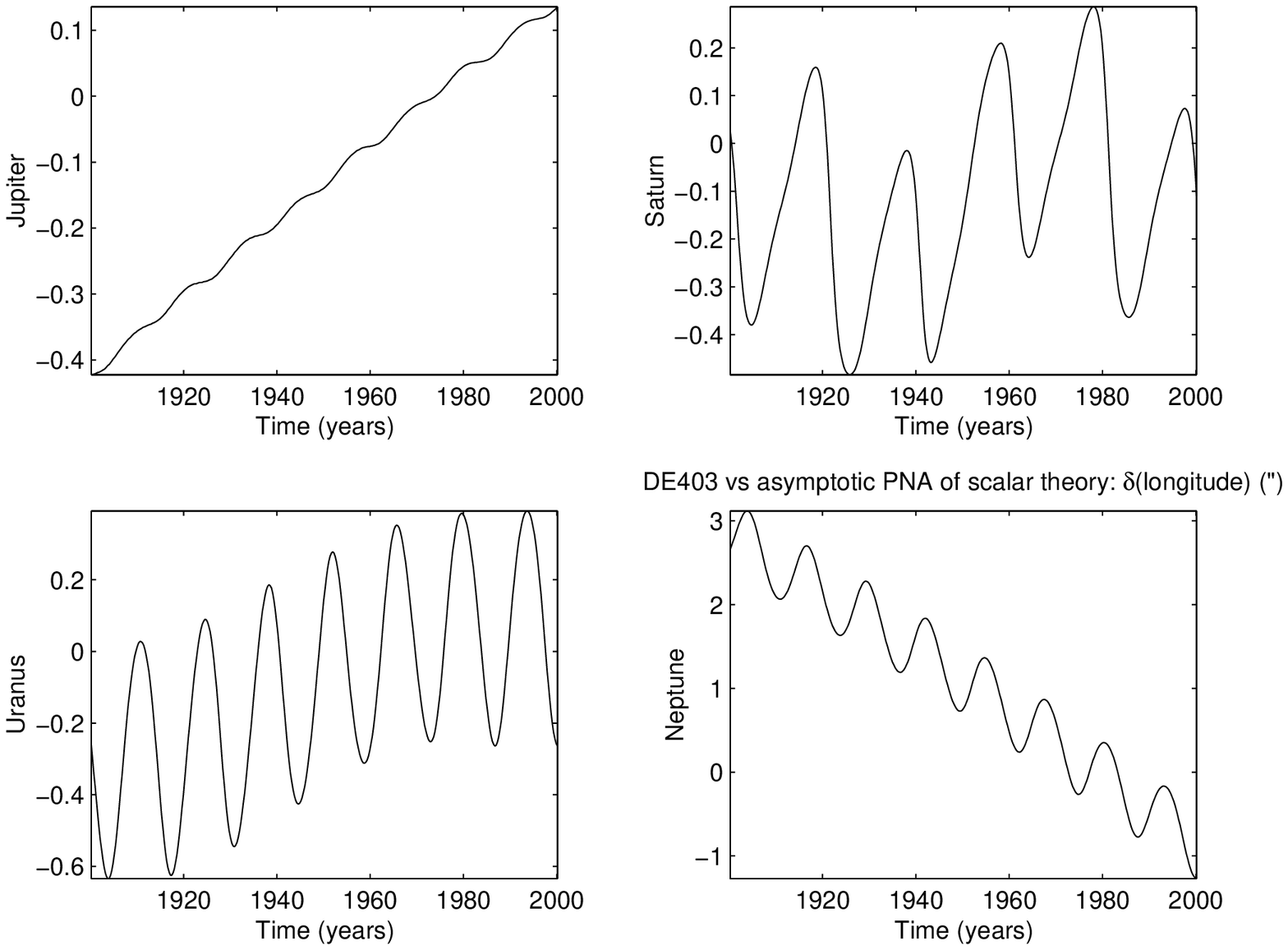}
	\caption{\small{Longitude differences (arc seconds) for 
	planets Jupiter to Neptune, either as taken from the DE403 ephemeris of the Jet
	Propulsion Laboratory, or as obtained (after adjustment) by
	numerical integration of the equations of motion in the asymptotic
	post-Newtonian approximation of the investigated scalar theory.}}
\end{figure} 

With these input data, themselves a fitting of observations by the
LDEIH equations of GR, the magnitude of the optimal vector is
$\abs{\mathbf{V}}\approx 3\mathrm{km/s}$, which is significant.
The difference between DE403 and our thus-adjusted equations of
motion is shown on figures 1 and 2. {\it The self-rotation of
all nine bodies is neglected}, {\it i.e.}, all $\boldsymbol{\omega}_a$ 's
are assumed $\mathbf{0}$ in (\ref{u-rigid}). We show only the
difference in longitudes, because it dominates over the other
errors. It can be seen that, for most planets, the difference is
very small (a few times $0.1''$) over the fitting interval, but it
increases quickly with time for the inner planets (Mercury, Venus,
the Earth-Moon Barycenter (EMB), and Mars). We do not know yet
whether this comes only from the different models or partly also
from numerical reasons: due to the necessity of reinitializing
very often (every two days here---we mean the ephemeris time, not
the computer time), the calculations are long. It has already been
checked that an increase of the accuracy in the ODE-solver brings
negligible changes to the present Figure showing the differences
between DE403 and the asymptotic PNA of the scalar theory. Anyhow,
the differences are still quite small, {\it e.g.} $3.8''$ for Mercury
after the last century, to be compared with the relativistic
perihelion advance of $43''$, and with the accuracy of the current
ephemerides, considered to be $0.1''$ for Mercury
\cite{Simon98},p. 228. For the influence of the Moon on the motion
of the EMB, we use a semi-analytical correction formula
\cite{Bretagnon80}, which we adapted in an approximate way from
the ``1950 ecliptic" to the ``J2000" reference.

\section{Conclusion}

The usual method of asymptotic expansions, as defined in Sect. 2,
is indeed of usual utilization in most domains where partial
differential equations occur, but it contrasts with the standard
(Fock-Chandrasekhar) method of post-Newtonian approximation (PNA) for weak
gravitational fields. In the latter method, no one-parameter
family of similar problems is introduced, so that the meaning of
the approximation is not very clear. Indeed $1/c^2$ is formally
considered as a small parameter, and the matter fields are not
expanded. We applied the usual method of asymptotic expansions to
weak gravitational fields in our scalar theory, and we call the
result the asymptotic PNA. A similar method has been proposed in
GR by Synge and coworkers \cite{Das61, Florides62} for stationary gravitational
fields, and has been initiated in a more general case, but not
fully developed, by Futamase \& Schutz \cite{FutaSchutz}. In our opinion, the
asymptotic PNA is very solid mathematically. Within the asymptotic
PNA, $1/c^2$ turns out to be (proportional to) the small parameter
$\lambda$, but this is true in specific units, depending on
$\lambda$. The asymptotic PNA leads to definitely different
equations, as compared with the standard PNA. In particular, the
former method predicts that the internal structure of the bodies,
and their internal motion, has a definite influence on the motion
of the mass centers of a self-gravitating system of bodies.\\

This has been checked numerically in the solar system for the
scalar theory. The standard PNA should probably lead to the same result in
the scalar theory as in GR, namely it should lead to the conclusion
that, in the solar system, the post-Newtonian effects may be
calculated simply by adding to the Newtonian motion the PN
corrections obtained in considering all planets as test particles
in the field of the isolated Sun. If the latter approximation is
used, the results of our scalar theory are nearly
indistinguishable from those of GR. \footnote{~{\it We did check
this numerically}. To this end, for each planet, we took the PN
equation of motion of a test particle \cite{Arm98b} (p. 22), the
first-order contribution assuming a spherical Sun. The equation of
motion then reduces to the ``Schwarzschild-corrected Newtonian
equation", plus extra terms which vanish if the velocity of the
Sun through the preferred frame is zero. As a result of fitting
these equations to the DE403 ephemeris, that velocity was found
negligible, and of course its effect on the thus-calculated
ephemeris was then found negligible also.} In contrast, if the
equations of the asymptotic PNA are used, then the predicted
motion apparently cannot fit a standard ephemeris ({\it i.e.} an
ephemeris based on the standard PNA of GR) within what is
currently believed to be the observational accuracy. \\

It seems that the influence of the internal structure of the bodies and the difference with the DE403 ephemeris are much more the result of changing the approximation
method, than that of changing the theory. Indeed, the exact local equations of motion for a perfect fluid in the scalar theory are very similar to those in GR, and the general metric of the theory is a ``Schwarzschild-like'' metric. In other words, the author
considers it likely that a similar departure from a standard
ephemeris would be left, if one compared it with a calculation
based on an asymptotic PNA of GR. However, this could be proved only by building a general asymptotic scheme in GR. This would be difficult, due to the fact that in GR the initial conditions cannot be imposed freely. \\

Coming back to the scalar theory, there is still a possibility to improve the fitting, mainly by
re-adjusting the masses (this could not be done here, except for
those of the Sun and Jupiter, because the masses are sensible
parameters whose values have to be separately optimized before
possibly optimizing them globally), perhaps also by improving the
numerical accuracy. It is also very important to adjust the
equations, not on an ephemeris (because this is already a fitting
of observations by some other equations), but directly on
observations. Indeed some correction factors of observational data
are taken as free parameters in the adjustment of an ephemeris,
hence the observations are not completely independent of the
gravitational model. Finally, it should be noted that the
(best-fitting) value of the absolute velocity $\mathbf{V}$ of the
mass center of the solar system has been found to be ca.~3 km/s
with the present model (self-rotation neglected, adjustment on a
standard ephemeris), and this already is not negligible. Due to
these simplifications, and due to the fact that the solar system
was assumed isolated here, this present best-fitting value of
$\mathbf{V}$ is not even an approximation to the correct one: it
justs tells an idea about the order of magnitude of $\mathbf{V}$.
That this is not negligible, might incline one to think that the
preferred-frame character of the theory is not redhibitory.

\bibliographystyle{amsplain}

\end{document}